\documentclass[aip,BMF,amsmath,amssymb,showpacs,floatfix,citeautoscript]{revtex4-1}

\usepackage{amsmath}
\usepackage{amssymb}
\usepackage{wasysym}
\usepackage{pifont}
\usepackage{graphicx,overpic} 
\usepackage{mathrsfs}
\usepackage{upgreek}

\newcommand*{\ellp}{\ell_{\rm P}}

\newcommand*{\ellh}{x_{\rm H}}
\newcommand{\ve}[1]{\ensuremath{\mbox{\boldmath$#1$}}}

\begin{document}
\title{Distribution of label spacings for genome mapping in nanochannels}
\author{D. \"Odman}
\affiliation{Department of Physics, University of Gothenburg, 41296 Gothenburg, Sweden}
\author{E. Werner}
\affiliation{Department of Physics, University of Gothenburg, 41296 Gothenburg, Sweden}
\author{K. D. Dorfman}
\affiliation{Department of Chemical Engineering and Materials
Science, University of Minnesota, Minneapolis, MN 55455, USA}
\author{C. R. Doering}
\affiliation{Center for the Study of Complex Systems, University of Michigan, Ann Arbor, MI 48109-1042, USA}
\author{B. Mehlig}
\affiliation{Department of Physics, University of Gothenburg, 41296 Gothenburg, Sweden}
\date{\today}
 
\begin{abstract}
In genome mapping experiments, long DNA molecules are stretched by confining them to very narrow channels, so that the locations of sequence-specific fluorescent labels along the channel axis provide large-scale genomic information.  It is difficult, however, to make the channels narrow enough so that the DNA molecule is fully stretched. In practice its conformations may form hairpins that change the spacings between internal segments of the DNA molecule, and thus the label locations along the channel axis. Here we describe a theory for the distribution of label spacings that explains the heavy tails observed in distributions of label spacings in genome mapping experiments.  
\end{abstract}

 \maketitle

\section{Introduction}
Long DNA molecules in ionic solution adopt random conformations.
In equilibrium, the size of such DNA blobs is determined by a balance between entropic forces and excluded-volume interactions \cite{Wang2017}. In order to experimentally study local properties (melting patterns, DNA-protein reactions, DNA-sequence information), it has been suggested to stretch the DNA molecule, either
by applying a force to both ends of the molecule 
\cite{Strick2000,Bustamante2000}, or by confining it to a nanochannel \cite{Wang2005,Metzler2006,Zhang2009,Werner2014}. 

In next-generation genomics, for example, the locations along the channel axis of sequence-specific
fluorescent labels attached to the DNA molecule can be read by microscopy \cite{Jo2007,Das2010,Lam2012,Michaeli2012,Kounovsky2017}, providing a genetic fingerprint. This genome mapping technique requires the molecule to assume an effectively linear conformation, so that its global extension is close to its contour length. This can be achieved by making the channel very narrow, of the order of the persistence length $\ellp$ of the DNA molecule or smaller \cite{Odijk1983,Huh2007,Odijk2008,Kim2011,Menard2013,Zhang2013}. 
However, if the channel is not narrow enough, turns (\lq hairpin\rq{} bends) reduce the DNA extension along the channel axis \cite{Odijk2006,Odijk2008,Cifra2009,Su11,Dai2012,Cifra2012,Chen2013BMF,Chen2014,Shin2015,Muralidhar2014,Chen2017}. In genome mapping experiments, such hairpins may cause errors by changing the spacings $X$ between the fluorescent labels. Larger hairpins may even change their order. 

The distribution of label spacings has been measured in experiments \cite{Reinhart2015,Sheats2015}. It
exhibits a heavy tail at small spacings. 
The origin of this tail is not understood; it could be due to hairpins or small deflections from locally straight conformations \cite{Reinhart2015}. There is neither a theory for this distribution, nor are there simulation results that quantify the microscopic DNA conformations.  Even the most efficient algorithms 
simulating steady-state conformations of discrete wormlike chains \cite{Tree2013a} have not reliably computed the tails of the distribution for such narrow channels. 

Here we derive a theory for the distribution of label spacings in narrow channels with channel widths
 of the order of the persistence length. We model the correlated process of hairpin bends by
the telegraph model introduced in Ref.~\citenum{Wer17}, and account for deflections from straight polymer segments using 
a Gaussian approximation \cite{burkhardt2010}. We find a closed expression for the distribution in the limit where hairpin distances are large. The theory explains why the distribution is significantly skewed: the heavy left tail is caused by many relatively short $S$-shaped hairpins, which do not arise due to the cooperative mechanism proposed elsewhere \cite{Dai2012}.  
The right tail is approximately Gaussian. The theory predicts that
the distribution depends sensitively on the channel width $D$.
Our results are in good agreement with measured label-spacing distributions for strongly confined DNA \cite{Reinhart2015} for wide channels ($D=50$ nm), and the theory says that hairpins are frequent. 
For narrower channels ($D=41$ nm) the theory predicts that hairpins are rare.  In the experiment, short label spacings are 
somewhat more frequently observed than predicted. We discuss reasons for this discrepancy. 

\section{Telegraph model}
To compute the distribution of label spacings along the channel axis, 
we use the model derived in Ref.~\citenum{Wer17}, projecting the three-dimensional DNA configurations $\ve x(t)$ onto the channel axis $x$. The model consists
of two parts: an ideal correlated random walk, and a bias that takes into account self avoidance. Note that this model is distinctly different from the accelerated-particle method.\cite{Burkhardt1997,Bicout2001} This method does not
take into account self avoidance. It applies therefore when there are no hairpins, that is in the extreme Odijk limit.

\begin{figure}
\begin{overpic}[width=0.6\columnwidth]{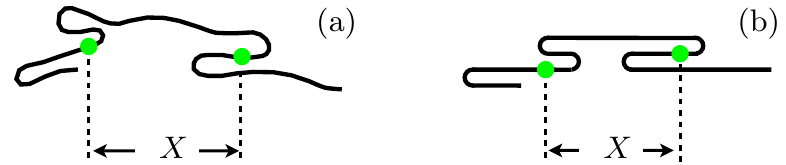}
\end{overpic}
\caption{\label{fig:telegraph} (a) Conformation of a confined DNA molecule. The distance along the channel axis
between the two fluorescent labels (green) is denoted by $X$. The conformation shown exhibits two S-hairpins and a C-hairpin at the left end. (b) Representation within the telegraph model. }
\end{figure}

We write the probability $P_1[x(t)]$ of
observing the one-dimensional, projected configuration $x(t)$ as
\begin{equation}
\label{eq:P}
P_1[x(t)] = P_{0}[x(t)]\, \mathscr{A}[x(t)]\,.
\end{equation}
Here $P_{0}$ is the distribution of the ideal telegraph process describing the position $x(t)$ at time $t$ of a particle moving with speed $v_0$. The particle changes its velocity $\pm v_0$ randomly at rate $r$. Random changes in the sign of the velocity give rise to $S$-hairpins and $C$-hairpins (Fig.~\ref{fig:telegraph}). The process lasts from $t=0$ to $t=T$. The parameters of the telegraph process map to those of the three-dimensional problem 
by letting $T$ be the contour length of the confined polymer and determining the parameters  $v_0$ and $r$ by comparing correlations.  The velocity correlations of the telegraph model decay exponentially \cite{Bal88} $\langle v(t) v(0) \rangle = v_0^2 \exp(-2rt)$.  The tangent correlations of the confined polymer have the same form \cite{Wer17}:
$\langle v_x(t) v_x(0)\rangle =a^2 \exp(-t/g)$, where the contour-length coordinate $t$ corresponds to time in the telegraph process. 
Further, $v_x$ is the $x$-component of the tangent vector $\ve v$ at $t$,
the parameter $a$ characterises the degree to which the tangent vectors align with the channel direction \cite{Werner2012}, and $g$ is the global persistence length \cite{Odijk2006}. 
One concludes that $v_0 = a$ and $r = (2g)^{-1}$.

The factor $\mathscr{A}$ in Eq.~(\ref{eq:P}) equals the fraction of three-dimensional polymer configurations that contain no colliding segments.
For narrow channels, an expression for $\mathscr{A}$  was derived in  Ref.~\citenum{Wer17}:
\begin{equation}
\label{eq:A}
\mathscr{A}[x(t)] = \exp\big[-\int{\rm d}x \,p_{\rm coll}(x)\big]\quad\mbox{with}\quad p_{\rm coll} =\frac{\varepsilon}{2v_0^2} N_{\rm s}(x)[N_{\rm s}(x)-1]\,.
\end{equation}
Here $N_{\rm s}(x)$ is the number of strands at location $x$, and
$\varepsilon$ parametrises the penalty for overlaps of the process.
This parameter depends on the persistence length $\ellp$ and  upon the effective width $w$ of the confined DNA molecule, as well as on the channel width $D$.
Assuming that a segment of length $\lambda= (\ellp D^2)^{1/3}$ (the Odijk deflection length\cite{Odijk1983}) is unlikely to overlap with another segment, an expression for $\varepsilon$ was derived in Ref.~\citenum{Wer17}:
\begin{equation}
\label{eq:penalty}
\varepsilon = \langle \delta(y - y')\delta(z - z') v_{\rm ex}\rangle_{\rm ideal}/\ell^2\,.
\end{equation}
Here $y$ and $z$ are transverse channel coordinates of a polymer segment of length $\ell$.  Primed coordinates label a second, independent segment,
and $v_{\rm ex}$ denotes the excluded volume between these two segments.
It depends upon the orientation of the segments. If $\ell \gg w$,
then $v_{\rm ex} = 2w\ell^2\sin\theta$, where $\theta$ is the angle between the two segments  \cite{Onsager1949}. The average $\langle \cdots\rangle_{\rm ideal}$ is over the conformations of the confined ideal polymer.

We can write the integral over the collision probability as
\begin{equation}
\int{\rm d}x \,p_{\rm coll}(x) = \frac{\varepsilon X}{v_0^2}\sum_{k=2}^\infty
\rho_k^{(0)} \tfrac{k(k-1)}{2}\,.
\end{equation}
Here $\rho_k^{(0)}$ is the fraction of channel positions $x$ with $k$ ideal DNA strands.
The $k\!=\!1$-term does not contribute to the sum.
Putting these results together we obtain:
\begin{equation}
\label{eq:Asum}
\mathscr{A}[x(t)] = \exp\big[- \tfrac{\varepsilon X}{v_0^2}\sum_{k=2}^\infty
 \tfrac{k(k-1)}{2}\rho_k^{(0)}\big]\,.
\end{equation}
In summary, Eq.~(\ref{eq:Asum}) describes conformation fluctuations on contour-length scales larger than  
the Odijk deflection length $\lambda$, and it is assumed that self avoidance
is sufficiently weak so that a hairpin of contour length $\lambda$  is unlikely to involve collisions. 
The bias $\mathscr{A}$ leads to a penalty against configurations of the ideal process with significant overlaps. In this model the effect of self avoidance is determined by the dimensionless parameter combination 
\begin{equation}
\label{eq:alpha}
\alpha = 
\frac{\varepsilon}{2r\,v_0}= \frac{\varepsilon g}{a}\,.
\end{equation}

In Ref.~\citenum{Wer17} the mean and variance of DNA extension in a nanochannel was computed numerically as a function of $\alpha$ by simulations of the telegraph model. It was shown that the results of direct numerical simulations of DNA extension in a nanochannel fall on universal scaling curves when plotted as a function of $\alpha$, as predicted by the telegraph model. The small-$\alpha$ behaviour of the {\em mean} extension and its {\em variance} was analysed in detail, and it was also argued that the extension variance scales as $\propto \alpha^{-3}$ for large $\alpha$, in the telegraph model.

In the following we show how to compute the {\em distribution} of label spacings in the limit of large $\alpha$ and large
$rT$. We obtain this distribution in closed form, consistent with the prediction that the variance is proportional to
$\alpha^{-3}$, and it allows to calculate the prefactor.

\section{Distribution of label  spacings for $rT\gg 1$ and $\alpha \gg 1$}
\subsection{Ideal process}

Consider first the ideal process. We can disregard the DNA-contour pieces to the left and to the right of the labels in Fig.~\ref{fig:telegraph}.
Assume that the process starts at $x(0)=0$ with $\pm v_0$ with equal probability. 
In the limit of $T\rightarrow\infty$ for fixed  $X'=X/(v_0T)$ the distribution $P_0(X,T)$  of $X= x(T)$ is derived in Appendix \ref{app:A}:
\begin{equation}
\label{eq:asymptP}
P_0(X',T) \sim   \frac{1}{2}\sqrt{\frac{rT}{2\pi}} \frac{1+\sqrt{1-X'^2}}{(1-X'^2)^{3/4}} {\rm e}^{rT(\sqrt{1-X'^2}-1)} \,,
\end{equation}
normalised to unity  on  $-1\leq X'\leq 1$,  for large $rT$ .  

\begin{figure}[t]
 \begin{overpic}[angle=0,width=0.6\columnwidth]{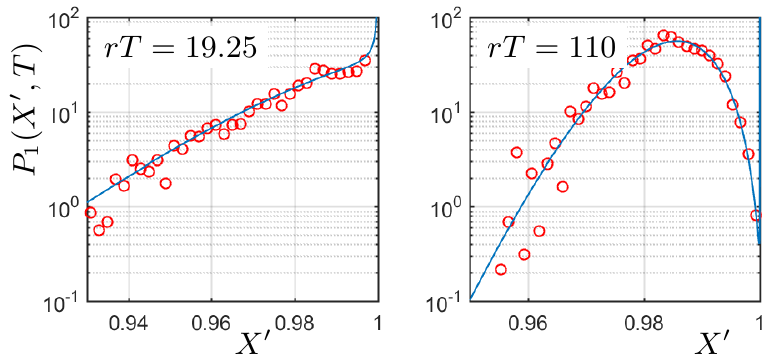}
 \end{overpic}
\caption{\label{fig:Dist_Telegraph}
  Distribution $P_1(X',T)$ of dimensionless labels spacings $X'$ along the channel axis, for the telegraph model. The blue lines show the theory, Eq.~(\ref{eq:18}). Symbols show  data from simulations of the telegraph model (see Appendix \ref{app:B}). Parameters left panel: $r=0.00275$, $T=7000$, $\varepsilon = 0.01$. Right panel: $r=0.0055$, $T=20000$, $\varepsilon = 0.02$.  Both panels: $v_0=1$ so that $\alpha=1.81$.}
 \end{figure}

\subsection{Self avoidance}
To take into account self avoidance for $\alpha\gg 1$, it suffices to consider the large-$X$ tail of $P_0(X,T)$, where $X\approx v_0T$. In this tail, hairpins 
are short and rare, so that they do not overlap. 
In the large-$T$ limit, the contribution of $C$-hairpins 
to $X$ is negligible in the ideal process,
because it is $O(1)$, whereas the contribution of $S$-hairpins
is $O(T)$.  We can therefore approximate  Eq.~(\ref{eq:Asum}) as 
$\mathscr{A}[x(t)] = \exp(3\rho_3^{(0)}{\varepsilon X}/{v_0^2} )$.
Normalisation implies: $\rho_1^{(0)}+\rho_3^{(0)}=1$. Contour length and extension
are related by $v_0T = X(\rho_1^{(0)}+3\rho_3^{(0)})$. 
Solving for $\rho_3^{(0)}$ yields 
\begin{equation}
\label{eq:Arho3}
\mathscr{A}[x(t)] = \exp\big[ \tfrac{3\varepsilon}{2v_0^2} (T v_0-X)\big]\,.
\end{equation}
Using Eqs.~(\ref{eq:P}), (\ref{eq:asymptP}) and (\ref{eq:Arho3}) 
we obtain an expression for the large-$X$ tail of the spacing distribution in the weakly self-avoiding telegraph model:
\begin{subequations}
\label{eq:18}
\begin{equation}
\label{eq:18a}
P_1(X',T) \sim  \mathscr{N} \tfrac{1+\sqrt{1-X'^2}}{(1-X'^2)^{3/4}}
{\rm e}^{-rT S(X')}\,,
\end{equation}
with normalisation factor $\mathscr{N}$, and with
\begin{equation}
\label{eq:18b}
 S(X') = 3\alpha(1-X')+ 1-\sqrt{1-X'^2}\,.
 \end{equation}
 \end{subequations}
 The distribution 
has a heavy left tail resulting from conformations  shortened by hairpins. 
 It depends on two dimensionless parameters, $\alpha$ and  $rT$.
The maximum of $P_1(X',T)$ is at 
$X'\sim 1-1/(18\alpha^2)$. 
Expanding the action $S(X')$ around this point we find 
$S(X') \sim S_{\rm min}+\tfrac{27}{2}\alpha^3\,\delta X'^2$.
This determines the variance:
\begin{equation}
\label{eq:result}
\sigma_1^2\,\tfrac{2r}{v_0^2 T} \sim \tfrac{2}{27}\alpha^{-3}\,.
\end{equation}
The numerical prefactor $\approx 0.09$ found  in Ref.~\citenum{Wer17} is
in reasonable agreement with $2/27 \approx 0.074$. 
Fig.~\ref{fig:Dist_Telegraph} shows results for $P_1(X,T)$ from computer simulations of the telegraph model
(see Appendix \ref{app:B}). We observe 
excellent agreement.

The model predicts microscopic conformation properties, such as the distribution
of hairpin lengths $\ellh$ along the channel axis. In the ideal process, hairpin turns are Poisson distributed
with rate $r$. The probability for two strands to overlap for 
a length $\ellh$ is $\propto\exp(- 2r \alpha\ellh/v_0)$, and there are three pairs of strands
to check for overlaps. So the probability of surviving the self-avoidance check is $\propto \mathscr{A}\sim \exp(-6r \alpha \ellh/v_0)$.
Thus the distribution of $\ellh$ is exponential with rate $6r \alpha/v_0$
\begin{equation}
\label{eq:pxh}
P(x_{\rm H}) = (6r \alpha/v_0) \,{\rm exp}(-6r \alpha x_{\rm H}/v_0)\,,
\end{equation}
and the mean hairpin length is
$\langle \ellh \rangle = {v_0}/{(6 r \alpha)} $.
In the limit $D/\ellp\to 0$, the global persistence length $g=1/(2r)$ 
diverges as $D/\ellp\to 0$ \cite{Wer17}. But since $r$ cancels out
in $\langle \ellh \rangle$ [see Eq.~(\ref{eq:alpha})], the mean hairpin length $\langle \ellh\rangle$ depends only weakly on 
$D/\ellp$. We conclude that overlaps become rare 
as $D/\ellp\to 0$ because there are fewer and
fewer hairpins, not because their length tends to zero.

\section{Comparison with experiment}
\citet{Reinhart2015} report experimental measurements of distributions of fluorescent label spacings. We now show that Eq.~(\ref{eq:18}) explains the shape of the measured distributions, and that it reveals the mechanism causing the substantial skewness of the measured distributions.
Up to this point we have neglected the effect of small deflections from straight contours. The telegraph model
takes into account the fact that the DNA segments need not align with the channel direction, but neglects the effect of small deflections on the extension. In narrow channels ($D\ll \ellp$),  Odijk's theory says that deflections cause the extension to slightly contract, and that the fluctuations of the extension around its mean are Gaussian in the limit of large  contour-length separations ($L\gg \lambda$),
with variance \cite{burkhardt2010}
\begin{equation}
\label{eq:Odijk}
 \sigma_{\rm Odijk}^2 = 0.0096\, L D_{\rm eff}^2/\ellp\,.
\end{equation}
Here and in the following, we express all results in terms of the physical variables contour length $L$, alignment factor $a$, and global persistence length $g$, instead of $T$, $v_0$, and $r$. For wider channels, the variance is expected to be larger than that given by Eq.~(\ref{eq:Odijk}), but a central-limit argument shows that the distribution remains Gaussian. 

We model the extension fluctuations  due to small deflections as Gaussian. Since the product of Gaussians is Gaussian we find for the label-spacing distribution:
\begin{align}
\label{eq:exp}
\mathscr{P}(X,L) &= \int_0^{aL}\! {\rm d} X_1 \int {\rm d}{\delta X} \,\,\delta(X-X_1+\delta X)
 \,P_1(X_1,L)\, \rho(\delta X)\\
 &= \int_0^{a L}\!\! {\rm d} X_1 P_1(X_1,L) \rho(X_1-X)\,.
 \end{align}
 Here $P_1(X,L)$ is the distribution of spacings $X$ between fluorescent labels with contour-length distance $L$ as obtained from the telegraph model,
 and
 \begin{equation}
 \label{eq:sigma0}
 \rho(\delta X) = (2\pi\sigma_0^2)^{-1/2} \exp[-\delta X^2/(2\sigma_0^2)]\,,
  \end{equation}
with free parameter  $\sigma_0^2$, the variance 
 due to small deflections (no hairpins). Eq.~(\ref{eq:sigma0}) gives 
$\sigma^2 = \sigma_{1}^2+\sigma_0^2$ for the extension variance,
where $\sigma_{1}^2$ is the extension variance in the telegraph model, asymptotic
to Eq.~(\ref{eq:result}). 
The label spacing distribution $\mathscr{P}(X,L)$ depends on three dimensionless parameters:
$\alpha$ (self avoidance), $L/(2g)$ (number of hairpin turns), and $\sigma_0/(aL)$ (effect of deflections).

\begin{table}[b]
\caption{\label{tab:params} Parameters for comparison with experiments in Fig.~\ref{fig:Reinhart}. $D_{\rm eff} = D - w$ is the effective channel width (see main text). Other parameters: $\ellp$ is the persistence length  of the DNA molecule,
$w$ is its effective width, $g$ is the global persistence length, $a$ is the alignment factor in the telegraph model.
The parameters $\ellp$, $w$, $g$, and $a$  are obtained from the Supplemental Material from Ref.~\cite{Wer17}
The parameters $\alpha$
and $\sigma_0$ are defined in Eqs.~(3) and (11) in the text.}
\begin{tabular}{lllllllll} \hline \hline
$D$ (nm) & $D_{\rm eff}$ (nm) & $\ellp$ (nm) & $w$ (nm) &  $g$ (nm) & $a$ & $\alpha$ &
$\sigma_0(28125\,{\rm bp})$ (nm) & $\sigma_0(53125 \,{\rm bp})$ (nm)\\ \hline
40 & 34.4 & 52    & 5.6 & 940 & 0.85\hspace*{4mm} & 8.53 \hspace*{4mm}& 72  & 117\\
42 & 36.4 & 52    & 5.6 & 743 & 0.84 & 6.13 & 72  & 117\\
51 & 45.4 & 52    & 5.6 & 362 & 0.81 & 2.10 & 112 & 176\\
53 & 47.4 & 52    & 5.6 & 321 & 0.80 & 1.74 & 99  & 137 \\
\hline\hline
\end{tabular}
\end{table}

\begin{figure}
\begin{overpic}[width=0.6\columnwidth]{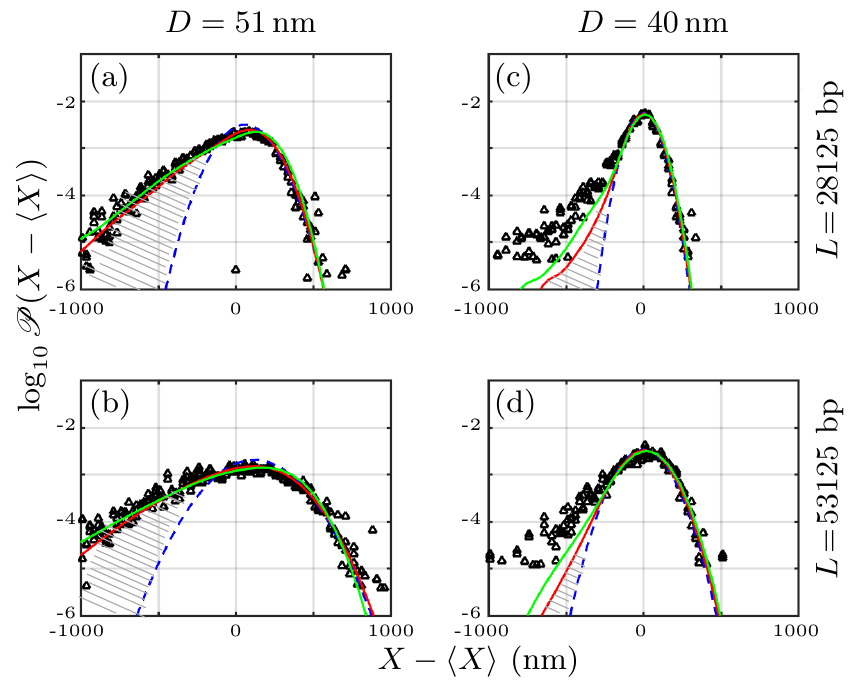}
 \end{overpic}
\caption{\label{fig:Reinhart} Comparison between the experimental data from Reinhart {\em et al.} (2015) \cite{Reinhart2015},  triangles, and theory, solid lines.
The theory is based on Eq.~(\ref{eq:exp}) and on  simulations of the telegraph model for $P_1(X,L)$. 
The hashed region indicates the contribution of hairpin-conformations. The solid red lines correspond to the effective channel widths  $D_{\rm eff}$ quoted in Table~\ref{tab:params}. The solid green lines were obtained using  $D_{\rm eff}$ that are $2$ nm larger (see text). The blue dashed line shows the Gaussian $\rho(\delta X)$, with $\sigma_0$ fitted (see text). }
\end{figure}

The parameter $\sigma_0$ in Eq.~(\ref{eq:sigma0})  could in principle be obtained by computer simulations of short 3D confined ideal wormlike chains, discarding all conformations that have hairpins. 
Here we take a  different approach. 
For the experimental parameters given in Table~\ref{tab:params},
the distribution $P_1(X,L)$ looks qualitatively like the distributions
shown in the left panels of Fig.~\ref{fig:Dist_Telegraph}: $P_1(X,L)$ increases
monotonously as a function of $X$ until $aL$, and it is zero for $X>aL$. As a consequence, Eq.~(\ref{eq:exp}) predicts that the right tail of $\mathscr{P}(X,L)$  (for $aL < X < L$) is caused by deflections in the absence of hairpins.  We can therefore determine $\sigma_0$ by fitting the right tail of $\mathscr{P}(X,L)$ to the experimental data. The fitted values of $\sigma_0$ (Table~\ref{tab:params}) are somewhat larger than the Odijk prediction in Eq.~(\ref{eq:Odijk})
-- by a factor of two, roughly. Given the value of $\sigma_0$, Eq.~(\ref{eq:exp}) yields a parameter-free prediction for
the effect of hairpins upon the distribution $\mathscr{P}(X,L)$ of fluorescent label spacings $X$ at contour-length separation $L$. 

The result is shown in Fig.~\ref{fig:Reinhart}. The Gaussian right tails (blue dashed lines in Fig.~\ref{fig:Reinhart})
 are good  fits to the experimental data \cite{Reinhart2015} (triangles). Also shown is
 the theory for $\mathscr{P}$, based on Eqs.~({\ref{eq:exp}},\ref{eq:sigma0}) and telegraph simulations
 for $P_1(X,L)$, red solid lines. The simulation parameters were obtained using the mapping derived in Ref.~\citenum{Wer17}, see Table~\ref{tab:params}. The parameter values in  this table  are based on the estimate
 $D_{\rm eff} =D-w$ for the  \lq effective channel width\rq{}, argued to take into account screened electrostatic interactions between the DNA and the channel walls \cite{Wang2011,Reisner2012}. However, the quoted expression for $D_{\rm eff}$ is just a rough estimate.  We have therefore run a second set of telegraph simulations for slightly larger values of $D_{\rm eff}$, namely $36.4$~nm and $47.4$~nm (solid green lines).
The parameters used are also given in Table~\ref{tab:params}.
 
 Panels (a) and (b) in Fig.~\ref{fig:Reinhart} show the comparison for the widest channel measured, for contour-length separations $L=53125$~bp and $28125$~bp. We estimate that these values correspond to $L=$ 18 $\upmu$m and 9.6 $\upmu$m,  assuming that the low intercalation used in these experiments does not affect $L$ \cite{Chuang2017}.
 We observe excellent agreement between theory and experiment. The hashed
 regions indicate the difference between the Gaussian approximation and the full theory, it corresponds to the contributions of hairpins  shortening the conformations. We see that the effect is substantial. We observe also that the extension distributions in Fig.~\ref{fig:Reinhart}(a,b)  are relatively insensitive to the precise channel size; the telegraph results are almost the same for the two different values of $D_{\rm eff}$. 

\section{Discussion and Conclusions}
Are the heavy left tails in Fig.~\ref{fig:Reinhart}(a,b) caused by few long hairpins or by many small ones? Using
the parameters from Table~\ref{tab:params} we find  that the mean hairpin length is $\langle \ellh\rangle \approx 47$ nm. 
This means that the typical hairpins are quite short, typically much shorter than the 2500 bp (850 nm) lower bound for resolving  nearest-neighbor fluorescent labels in experiments \cite{Chuang2017}. This is an important observation because the experimental data were conditioned on the sequence of fluorescent labels \cite{Reinhart2015}. Conformations that did not agree with the order of labels obtained from a reference genome were discarded. If hairpins frequently changed the order of fluorescent labels, the conditioning would cause a bias in the extension distribution. Here we can conclude that this effect is likely to be small, because the hairpins are quite short.

 This also means that expected hairpin contour length is of the order of the deflection length $\lambda$, which implies that the assumptions of the theory are only marginally met for the data from Ref.~\citenum{Reinhart2015}. 

Now consider the narrow-channel data in Fig.~\ref{fig:Reinhart}(c,d). The right tail of the extension is approximately Gaussian (blue dashed lines), and the left tails are heavy (solid red lines). But the experiments give a larger probability of very small spacings than the theory. In other words,  the experimental distribution is more skewed than the theory predicts. The theory, in fact, predicts that the effect of hairpins is negligible for $D=40$  nm, as opposed to the $D= 51$ nm channel.  We do not know the reason for this discrepancy. It might be that the most important hairpins are too short
for our theory to apply, but this is not likely as $\langle \ellh\rangle$ is of the same order as in the wide channel. 
We cannot exclude that there are other reasons that bias the experimental  data  to smaller spacings. 
Using a different experimental method, \ \citet{Sheats2015} suggested that the distributions are somewhat less skewed in narrow channels.  However, we believe that the exposure time in the analysis by \ \citet{Sheats2015} was too short to allow for a reliable estimate of the tails of the distribution and too few
independent conformations were sampled. 

Comparing the theoretical predictions for slightly different channel widths ($D=40$~nm and $42$~nm) we infer that the theoretical results are quite sensitive to the value of the effective channel width -- which is not known precisely. The theory for $42$~nm channels is much closer to the experimental data. The sensitive dependence is of interest because it reflects the fact that hairpin formation is an activated process. 

What remains to be done? First, the Gaussian model for the right-hand tail of the spacing
distribution $\mathscr{P}(X,L)$ is highly idealised, and the parameter $\sigma_0$ was fitted to the data. The exact dependence of $\sigma_0$  upon $L$ could be obtained from simulations of confined discrete wormlike chains \cite{Tree2013a}, by measuring the fluctuations of the end-to-end distance of short chains conditional on no hairpins, or using propagator methods \cite{Chen2017}.
It may be possible, but much more difficult, to simulate the full distribution to determine how our theory breaks down as $\langle \ellh\rangle$ becomes smaller.  Second, the sensitive dependence of the predicted distribution upon channel size can test theories for the 
effective channel width \cite{Reisner2012}, an important open question for genome mapping experiments.
We further expect that the tails are sensitive to flexibility of the DNA backbone, 
providing a probe for sequence-specific effects on confined DNA conformations, an emerging area of interest \cite{Chuang2017}. 
Finally, an entirely open question is to understand the conformational dynamics \cite{levy_entropic_2008,ibanez-garcia_hairpin_2013,Alizadehheidari2015,Werner2018,Krog2018}. While the results described here show that the equilibrium conformation statistics of strongly confined DNA molecules are now well understood, 
much less is known about the dynamics.

\begin{acknowledgements}
This work was supported by VR grants 2013-3992 and 2017-3865, by NIH grant R01-HG006851, 
and by NSF grant DMS-1515161.
Computational resources were provided by  C3SE and SNIC.
\end{acknowledgements}


%
 
 \appendix
 
 \section{Ideal distribution}
\renewcommand\theequation{\arabic{equation}}
\addtocounter{equation}{15}
 \label{app:A}
\begin{figure}[b]
\begin{overpic}[width=0.4\columnwidth]{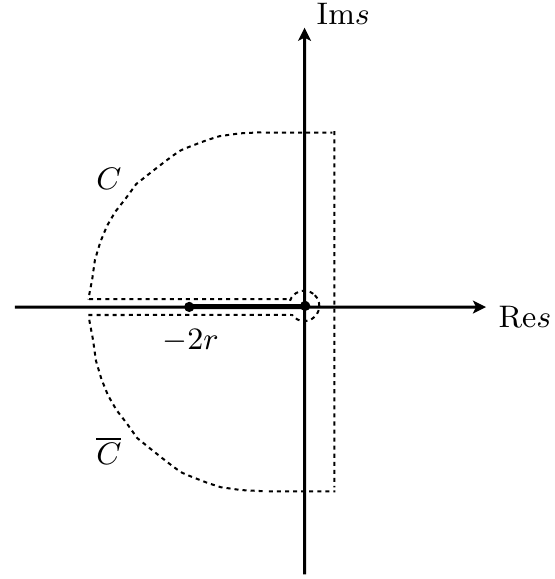}
\end{overpic}
\caption{\label{fig:contour}Contour for inverse Laplace transform for $0 < x < v_0t$.
There is a branch cut at $[-2r,0]$.}
\end{figure}

Assume that the polymer starts at $x=0$ at $t=0$ with $\pm v_0$ with equal probability.
Denote the distribution of $x(t)$ by $p(x,t)$. Decompose $p(x,t)=\rho_+(x,t)+\rho_-(x,t)$, where $\rho_+(x,t)$ is the probability to find the process at $x$ at time $t$ with velocity $v_0>0$, and $\rho_-(x,t)$ is the probability to find the process at $x$ at time $t$ with velocity $v_0<0$.
Then:
\begin{equation} \label{eq:ME}
\frac{\partial}{\partial t} \left[ \begin{array}{l} \rho_+(x,t)\\\rho_-(x,t)\end{array}\right]=
\left[ \begin{array}{ll} -r-v_0\partial_x & \phantom-r\\\phantom-r & -r + v_0 \partial_x\end{array}\right]
\left[ \begin{array}{l} \rho_+(x,t)\\\rho_-(x,t)\end{array}\right]\,.
\end{equation}
Let $q(x,t)= \rho_+(x,t)-\rho_-(x,t)$:
\begin{align}
\label{eq:ME2}
\partial_t p &= -v_0 \partial_x q\,,\quad\mbox{and}\quad \partial_t q    = -2r q - v_0 \partial_xp\,.
\end{align}
To solve (\ref{eq:ME2}) we use the Laplace transform. Denote the Laplace transforms $\mathscr{L}$ of $p$ and $q$ by $P= \mathscr{L}(p)$ and $Q= \mathscr{L}(q)$. Since the process starts at $x=0$ with $\pm v_0$
with equal probability, the  initial conditions are $q(x,0) = 0$ and $p(x,0) = \delta(x)$.
As a consequence the Laplace transforms obey:
\begin{equation}
\label{eq:sPQ}
sP(x,s)-\delta(x) = -v_0 \partial_xQ(x,s)\quad\mbox{and}\quad  sQ(x,s) = -2rQ(x,s)-v_0 \partial_xP(x,s)\,.
\end{equation}
The solution of Eq.~(\ref{eq:sPQ}) reads:
\begin{align}
Q(x,s) &= -\frac{sv_0}{\sigma^2} \frac{\rm d}{{\rm d}x} P(x,s)\quad\mbox{and}\quad
P(x,s) = C_1 {\rm e}^{\sigma x/v_0} + C_2 {\rm e}^{-\sigma x/v_0}
- H(x)\frac{\sigma}{2sv_0}  \big( {\rm e}^{\sigma x/v_0} - {\rm e}^{-\sigma x/v_0 }\big)\,,
\label{eq:PC1C2}
\end{align}
where $H(x)$ is the Heaviside function and $\sigma = \sqrt{s^2+2sr}$.
The boundary condition $p(-\infty,t)=0$ gives
$C_2 = 0$,
and  $p(\infty,t)=0$ yields
$C_1 = {\sigma}/({2sv_0})$.
Inserting these expressions for $C_1$ and $C_2$ into Eq.~(\ref{eq:PC1C2}) we find:
\begin{align}
P(x,s)&= \frac{\sigma}{2s v_0}
\big[H(x) {\rm e}^{-\sigma x/v_0} + H(-x) {\rm e}^{\sigma x/v_0}\big]
\quad\mbox{and}\quad Q(x,s) = \frac{1}{v_0}\big[ H(x) {\rm e}^{-\sigma x/v_0}-H(-x) {\rm e}^{\sigma x/v_0}\big]\,.
\end{align}
Now assume $x>0$ and evaluate the inverse Laplace transform
\begin{equation}
p(x,t) = \frac{1}{2v_0}\int_{\gamma-i\infty}^{\gamma+i\infty}\!\!
\frac{{\rm d}s}{2\pi i} \frac{\sigma}{s} {\rm e}^{-\sigma x/v_0+st}
\end{equation}
using contour integration. For $x>v_0t$, we close the contour in the right half plane.
This contour contains no singularities, so $P(x,t)=0$.
When $0<x<v_0t$, we close the contour in the left half plane as shown in
Fig.~\ref{fig:contour}. The integral along $C$ and $\overline{C}$ vanishes, the only contribution comes from integration around the branch cut:
\begin{equation}
\label{eq:Pint0}
\frac{1}{\pi}\!
\int_0^{2r}\!\!{\rm d}s\, \frac{\sqrt{2sr-s^2}}{sv_0}
\cos(\sqrt{2sr-s^2}x/v_0) {\rm e}^{-st}\,.
\end{equation}
When $x=v_0t$ then the integrals along  $C$ and $\overline{C}$
diverge. This gives a contribution proportional to
$\delta(x-v_0t)$ to $P(x,t)$.
Since Eq.~(\ref{eq:Pint0}) is normalised to $1-{\rm e}^{-rt}$ when integrating from $x=0$ to $x=v_0t$
we have:
\begin{equation}
\label{eq:Pint}
p(x,t) = {\rm e}^{-rt}\delta(x-v_0t) + H(x-v_0t) \,\,\frac{1}{\pi}\!
\int_0^{2r}\!\!{\rm d}s\, \frac{\sqrt{2sr-s^2}}{sv_0}
\cos(\sqrt{2sr-s^2}x/v_0) {\rm e}^{-st}\,,
\end{equation}
normalised on $[0^-,\infty]$.
Numerical evaluation of the smooth part of Eq.~(\ref{eq:Pint}) for $v_0=1$ and different
values of $r$ and $t$ shows that this result is equivalent to an expression in Ref.~\citep{Bal88}:
\begin{equation}
\label{eq:pxt2}
 p(x,t) = {\rm e}^{-rt} \delta(x-v_0 t) + H(x-v_0 t)\frac{r{\rm e}^{-rt}}{v_0} \Big[I_0(rt\sqrt{1-x^2/(v_0t)^2})
 + \frac{I_1(rt\sqrt{1-x^2/(v_0t)^2})}{\sqrt{1-x^2/(v_0t)^2}}  \Big]\,.
\end{equation}
In the limit of $T\rightarrow\infty$
for fixed $x'=x/(v_0T)$, stationary-phase evaluation of the integral
in Eq.~(\ref{eq:Pint}) yields
\begin{equation}
P_0(x',T) \sim   \frac{1}{2}\sqrt{\frac{rT}{2\pi}} \frac{1+\sqrt{1-x'^2}}{(1-x'^2)^{3/4}} {\rm e}^{rT(\sqrt{1-x'^2}-1)} \,.
\end{equation}
This is Eq.~(\ref{eq:asymptP})  in the main text, noting that the label spacing is given
by $X=x(T)$ since $x(0)=0$. We remark that Eq.~(6) is normalised to unity for $-1 \leq X' \leq 1$.

\section{Description of Monte Carlo algorithm for simulation of telegraph model with self avoidance}

\renewcommand\theequation{\arabic{equation}}
\addtocounter{equation}{25}

\label{app:B}
This Section describes the implementation of an algorithm to simulate the telegraph model with self avoidance.
The telegraph process is implemented as described in the Supplemental Material of Ref.~\cite{Wer17}, and the telegraph simulations use a modified version  \cite{dai2014b,Smithe2015} of the PERM  algorithm \cite{grassberger1997,Prellberg2004}.
The algorithm grows an ensemble of $N$ polymers, each represented by a
discrete telegraph process. Initial conditions: $x_0=0$ and $v_0 = -1$ or $+1$
with equal probability. The polymers grow in the
direction of $v_0$. After each time step $t$, each polymer has a chance $r$ of changing sign of $v_0$, so that it continues to grow in the opposite direction.

If the polymer reaches a site that it has already visited $n$ times before, it has a chance ${\rm e}^{-\epsilon n}$ of surviving to the next time step.
In the simplest form of the algorithm, the polymer is discarded if it does not
survive. Attrition renders this algorithm inefficient. Therefore we
used the modified version \cite{dai2014b,Smithe2015},
where $N_b$ batches each containing $N_{p/b}$ polymers are grown simultaneously
($N=N_b N_{p/b}$). Every time a polymer fails the survival check, it is replaced by a polymer randomly sampled from its batch.

The replacement of polymers creates a bias in the statistics conformation
statistics \cite{dai2014b}, which must be  corrected for. This can be done by introducing weights as described in Ref. \cite{dai2014b}.
Every polymer in a given batch is assigned
the same initial weight $w_0 = N_{p/b}^{-1}$.
After each time step the weights of all polymers in a given batch
are updated as $w_t = w_{t-1}\frac{N_s}{N_{p/b}}$, where $N_s$ is the
number of polymers in the batch that survived the check.
This rule decreases the weights of batches where many polymers have been replaced.

This process continues for $T$ time steps. For each polymer,
the extension $X_T$ along the channel axis is measured. Mean $\mu$ and variance $\sigma^2$ of the extension are calculated as
\begin{align}
\mu &= \frac{\sum_{k = 1}^{N} w^{(k)}_T X^{(k)}_{T}}{\sum_{k = 1}^{N} w^{(k)}_T}\\
\sigma^2 &=  \frac{\sum_{k = 1}^{N}w^{(k)}_T[X^{(k)}_{T}-\mu]^2}{\sum_{k = 1}^{N} w^{(k)}_T},
\end{align}
where $k$ labels all polymers, across all batches, and $w^{(k)}$ are the corresponding weights.

\end{document}